# Oblivious RAM Simulation with Efficient Worst-Case Access Overhead

Michael T. Goodrich  
Univ. California, Irvine  
goodrich@ics.uci.edu

Michael Mitzenmacher  
Harvard University  
michaelm@eecs.harvard.edu

Olga Ohrimenko  
Brown University  
olya@cs.brown.edu

Roberto Tamassia  
Brown University  
rt@cs.brown.edu

October 17, 2018

**Abstract**

Oblivious RAM simulation is a method for achieving confidentiality and privacy in cloud computing environments. It involves obscuring the access patterns to a remote storage so that the manager of that storage cannot infer information about its contents. Existing solutions typically involve small amortized overheads for achieving this goal, but nevertheless involve potentially huge variations in access times, depending on when they occur. In this paper, we show how to de-amortize oblivious RAM simulations, so that each access takes a worst-case bounded amount of time.

## 1 Introduction

In the cloud storage model, a user, Alice, remotely stores a large set of data with a remote server, Bob, offloading her need to maintain her data. To achieve confidentiality from Bob, Alice should store her data in encrypted form, but encryption alone is not enough, since information about her data may be leaked by the pattern in which she accesses it.

Oblivious RAM simulation tackles this privacy-protection problem by hiding data access patterns from Bob. Such solutions simulate a general random access machine (RAM) computation with respect to the external storage, but they perform additional obfuscating accesses that hide the locations of requests. Recently there has been a considerable amount of work on methods that optimize the *access overhead*, i.e., the number of additional accesses to the data repository per request to provide oblivious access (e.g., see [7, 6, 1, 12, 10, 2, 14, 5]). Nevertheless, existing oblivious RAM solutions still suffer from a worst-case access overhead that can be as bad as $\Omega(n)$ for $n$ data items. This worst-case overhead makes oblivious RAM unsuitable in many practical scenarios, such as real-time systems and multi-user systems. For example, consider a multi-user environment where a group of users share the same external storage: one of the users has to make $\Omega(n)$ accesses in the worst case to make sure all accesses remain oblivious. And she may have to spend this much time with each of her accesses. Thus, we are interested in this paper of *de-amortized* solutions that have the same access overhead for every request.

### 1.1 The Oblivious RAM Model

An *oblivious RAM* (*ORAM*) is an interface between a client, Alice, and a data repository, Bob, whereby Alice outsources the storage of $n$ data items to Bob. When Alice makes a *request* for item $x$, she issues a sequence of *accesses* to Bob's data repository to retrieve $x$ in such a way that Bob is unable to determine which item is being accessed (any better than a random guess). Of course, there is a simple way for Alice to obfuscate her requests—she could simply read all items from the data repository with each request (her requested item $x$ is sure to be in this collection). But such a solution has access overhead $\Theta(n)$; hence is quite inefficient.

Table 1: Comparison of Oblivious RAM simulation Methods.

|  | Client Memory | Server Storage Overhead | Amortized Access Overhead | Worst-Case Access Overhead |
|---|---|---|---|---|
| Goldreich-Ostrovsky [5] $\sqrt{n}$ | $O(1)$ | $O(n)$ | $O(\sqrt{n}\log^2 n)$ | $O(n\log^2 n)$ |
| Goldreich-Ostrovsky [5] $\log n$ | $O(1)$ | $O(n\log n)$ | $O(\log^3 n)$ | $O(n\log^2 n)$ |
| Williams *et al.* [14] | $O(\sqrt{n})$ | $O(n\log n)$ | $O(\log^2 n)$ | $O(n\log n)$ |
| Goodrich-Mitzenmacher [6] (1) | $O(1)$ | $O(n)$ | $O(\log^2 n)$ | $O(n)$ |
| Kushilevitz *et al.* [10] | $O(1)$ | $O(n)$ | $O(\log^2 n/\log\log n)$ | $O(n)$ |
| Stefanov *et al.* [12] | $O(\sqrt{n})$ | $O(n)$ | $O(\log^2 n)$ | $O(\sqrt{n})$ |
| Goodrich-Mitzenmacher [6] (2) | $O(n^\nu)$ | $O(n)$ | $O(\log n)$ | $O(n)$ |
| Goodrich *et al.* [7] | $O(n^\nu)$ | $O(n)$ | $O(\log n)$ | $O(n)$ |
| Boneh *et al.* [1] | $O(\sqrt{n\log n})$ | $O(n)$ | $O(1)$ | $O(n\log n)$ |
| **De-amortized $\log n$** | $\boldsymbol{O(n^\tau)}$ | $\boldsymbol{O(n)}$ | $\boldsymbol{O(\log n)}$ | $\boldsymbol{O(\log n)}$ |
| De-amortized $\sqrt{n}$ | $O(1)$ | $O(n)$ | $O(\sqrt{n}\log^2 n)$ | $O(\sqrt{n}\log^2 n)$ |

## 1.2 Related Prior Work

Prior work on oblivious RAM addresses the trade-off between the size of the client's memory, the access overhead, and the space overhead at the data repository, i.e., the additional space used beyond the $n$ items. Based on the assumptions about the client, oblivious RAM models can be split into *stateless* and *stateful* solutions. A stateless oblivious RAM is not allowed to keep a state between requests and hence can be used in a multi-user scenario. Stateful solutions assume Alice keeps information in a private storage (which she maintains), which helps her perform her accesses obliviously in the remote storage.

Stateless oblivious RAM simulation was first proposed by Goldreich and Ostrovsky in [5], who present a preliminary simple solution with $O(\sqrt{n}\log^2 n)$ amortized access overhead, referred as the *square-root solution*, and a more complex solution with $O(\log^3 n)$ amortized access overhead. Goodrich and Mitzenmacher [6] improve this result by giving a method with $O(\log^2 n)$ amortized access overhead with high probability. Recently Kushilevitz *et al.* [10] show that techniques from [6] can be extended to obtain $O(\log^2 n/\log\log n)$ amortized access overhead. All the above methods are stateless and consider a private memory of size $O(1)$ for Alice, an overly restrictive assumption in practice.

Other solutions [7, 13, 14] improve the overall access overhead by assuming that a client has a workspace of non-constant size. Williams and Sion [13] achieve $O(\log^2 n)$ expected amortized access overhead and $O(n\log n)$ space overhead with $O(\sqrt{n})$ private memory. Williams *et al.* [14] improve the method from [13] to achieve $O(\log n\log\log n)$ amortized access overhead. Goodrich *et al.* [7] give an oblivious RAM simulation method with $O(\log n)$ amortized access overhead given that a client has access to workspace of size $O(n^\nu)$, for a given constant $\nu > 0$.

Other recent papers provide stateful solutions, i.e., where a client maintains a state between requests to the data repository in a non-constant sized private cache. A RAM simulation by Goodrich and Mitzenmacher [6] achieves an overhead of $O(\log n)$ and uses a private cache of size $O(n^\nu)$, for any given fixed constant $\nu > 0$, which maintains a state. Boneh *et al.* [1] propose a scheme that achieves an amortized overhead of $O(1)$ but using a cache of size $O(\sqrt{n\log n})$, which also maintains state.

Damgård *et al.* [2] and Goodrich *et al.* [7] present stateless oblivious RAM simulations without cryptographic assumptions about the existence of random hash functions. Damgård *et al.* [2] show that amortized access overhead of $O(\log^3 n)$ is possible for oblivious RAM simulation without using random functions.

Goodrich *et al.* [7] present a method with $O(\log^2 n)$ amortized access overhead that also does not use random functions.

To sum up, then, all the methods described above have amortized access overheads. (See Table 1 for the comparison of ORAM simulations.) Indeed, even the most efficient previous solutions can incur an $O(n)$ overhead in the worst case for any given request. This is slightly improved in recent work by Stefanov *et al.* [12], who give an oblivious RAM simulation that achieves an $O(\sqrt{n})$ access overhead in the worst case, while having $O(\log^2 n)$ amortized overhead complexity; hence, their solution is also amortized, but not as inefficient as previous schemes on a per-access basis.

Oblivious RAM simulation has also been used to protect against traffic analysis in a networked file system [15].

Kosaraju and Pop [9] give an overview of general de-amortization techniques. E.g., one of the techniques, de-amortization via data duplication [3], maintains two copies of the data set: one for performing the redistribution of the data and one for accesses. We cannot apply these general techniques to our problem, however, since we also need to ensure the obliviousness of the de-amortized algorithm.

### 1.3 Our Results

We present two oblivious RAM simulations that achieve a sublinear access overhead on every request made by the client to the data repository. The first is a de-amortized version of the square root solution originally presented in [5]. This method has $O(\sqrt{n} \log^2 n)$ access overhead in the worst case while using $O(n)$ space on the data repository and assuming $O(1)$ workspace on the client side. We then de-amortize an efficient oblivious RAM simulation by Goodrich *et al.* [7], that we refer to as the "$\log n$ hierarchical" solution. In this solution, we achieve wost-case access overhead of $O(\log n)$ and space overhead of $O(n)$, assuming that a client has access to a workspace of size $O(n^\tau)$, for any given fixed constant $\tau > 0$.

## 2 Preliminaries

We assume that the client outsources $n$ data items to a remote data repository that supports the following *access operations*:

- $read(i)$: return the content of location $i$;
- $write(i, x)$: write data item $x$ to location $i$;
- $copy(i, j, count)$: copy a memory block of size $count$ from location $i$ to location $j$.

The latter block-copy operation is not actually required by our methods but using it makes the algorithms more intuitive.

We also assume that provider of the storage service, Bob, is an *honest-but-curious* adversary [4], in that he correctly performs all operations and does not tamper with the data.

A data item is stored by Alice in the repository as the encryption of a pair $(x, v)$, where $x$ is the virtual address of the item in the RAM and $v$ is its value. Typically, oblivious RAM solutions use *probabilistic encryption* to make sure that Bob cannot distinguish between reads and writes or track repeated accesses to the same data item. Namely, Alice encrypts each data item that she writes to the data repository using a probabilistic encryption scheme based on her private key. Also, Alice reencrypts and rewrites each data item she accesses so that its encryption will change even if the data item is not modified. This technique ensures that Bob is computationally unable to determine the plaintext of any memory cell from that cell's contents alone. Also, it is unfeasible for Bob to determine whether two memory cells store encryptions of the same data item.

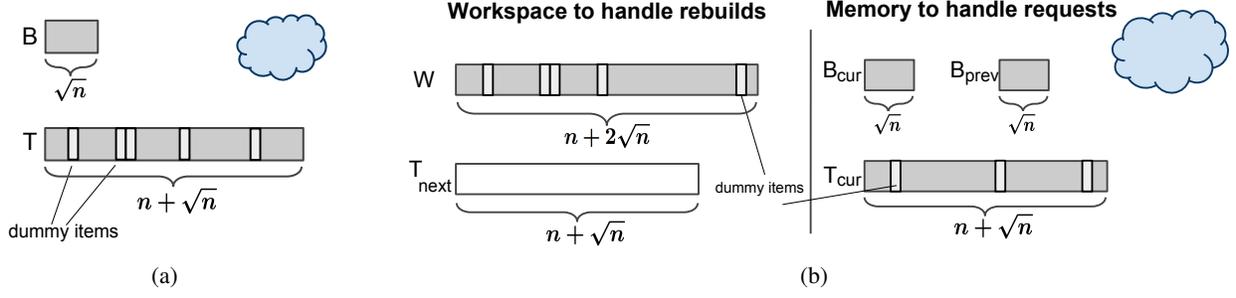

Figure 1: Memory layout of the data repository during oblivious RAM simulation: (a) original version of the square-root solution [5]; (b) our de-amortized version (Section 3.1).

## 2.1 Square-Root Solution

We first give an overview of the square-root oblivious RAM simulation method [5], which has $O(\sqrt{n}\log^2 n)$ amortized access overhead. We give enough details about the method for a reader to understand our de-amortized version provided in Section 3.1. Please refer to [5] for the full description.

The square-root solution uses storage space of size $n + 2\sqrt{n}$ at the data repository. (See Figure 1(a).) This space is split into a buffer, $B$, and a table, $T$. The buffer $B$ has size $\sqrt{n}$ and is used to cache the last $\sqrt{n}$ requests. Table $T$ contains a pseudo-random permutation of the $n$ data items and $\sqrt{n}$ dummy items.

Each data item is associated with a key (virtual RAM address) $x$, $x = 1, \cdots, n$ and each dummy item is given a key $n + d$ where $d = 1, \cdots, \sqrt{n}$. All items in $T$ are ordered according to a pseudo-random permutation function $\pi$ such that $\pi(x)$ gives the location of the item with key $x$ in $T$. (The full solution, which uses a binary search, is omitted in our description since the overall complexity of the request is $O(\sqrt{n})$.) The square-root ORAM simulation method is outlined in Algorithm 1. Note that table $T$ has to be rebuilt after every $\sqrt{n}$ requests. The rebuild phase consists of obliviously replacing the items in $T$ for which there is a new instance in $B$, associating the keys of real and dummy items with tags from a new permutation

---

**Algorithm 1** Oblivious RAM simulation using the square-root approach [5].

    Generate pseudo-random permutation function $\pi$
    Initialize table $T$ by storing the $n$ data items and $\sqrt{n}$ dummy items according to permutation $\pi$
    $count \leftarrow 1$
    **while** $true$ **do** {process a request}
      $found \leftarrow false$
      Scan all the locations in buffer $B$. During the scan, if data item $x$ is found, set $found \leftarrow true$.
      **if** $found$ **then**
        Access location $\pi(n + count)$ in $T$ {a dummy item}
      **else**
        Access location $\pi(x)$ in $T$ {data item $x$}
      **end if**
      Rewrite $B$, adding or replacing data item $x$
      $count \leftarrow count + 1$
      **if** $count > \sqrt{n}$ **then**
        Generate pseudo-random permutation function $\pi'$
        Construct a new table $T'$ with $\pi'$ using items in $T$ and $B$.
        clear $B$ and set $T \leftarrow T'$, $\pi \leftarrow \pi'$, and $count \leftarrow 1$
      **end if**
    **end while**

$\pi'$, and sorting items in $T$ according to $\pi'$. The rebuild phase takes $O(n \log^2 n)$ accesses. Since the rebuild happens only once every $\sqrt{n}$ requests, the amortized access overhead per request is $O(\sqrt{n} \log^2 n)$: $O(\sqrt{n})$ accesses for the request phase and $O(\sqrt{n} \log^2 n)$ accesses for the rebuild phase.

*Obliviousness:* Requests are handled by scanning buffer $B$ and accessing $T$. Due to the scheduled rebuilds, data items are associated with new tags every $\sqrt{n}$ requests. Between the rebuilds, unique locations are accessed in $T$: either an item $x$ is not present in the buffer and hence a unique location, $\pi(x)$, is accessed, or a unique dummy item, $\pi(n + count)$, is accessed.

## 2.2 $\log n$ Hierarchical Solution

We now describe the efficient oblivious RAM simulation by Goodrich *et al.* [7]. This method has $O(\log n)$ amortized access overhead, $O(n)$ space overhead at the repository, and client memory of size $O(n^\nu)$, where $\nu$ is an arbitrary positive constant. This method is stateless and thus suitable for a multi-user scenario since no client keeps a state between requests.

### 2.2.1 Cuckoo Hash Table

Since the method of [7] uses *cuckoo hash tables*, we give a short description of a cuckoo hash table. (See [11] for more details.).

A *cuckoo hash table* for $n$ items consists of two hash tables, $T_1$ and $T_2$ with two hash functions, $h_1$ and $h_2$. Each hash table contains $m = (1 + \epsilon)n$ memory cells, with $\epsilon$ a (small) positive constant. An item with key $x$ is found either in location $h_1(x)$ in $T_1$ or in $h_2(x)$ in $T_2$. We say that a failure occurs when there are only two possible locations for three keys $x, y, z$, i.e. $h_1(x) = h_1(y) = h_1(z)$ and $h_2(x) = h_2(y) = h_2(z)$. In this case a stash, $S$, of constant size is used to keep items that did not find a place in tables $T_1$ and $T_2$. The retrieval of an item now includes a scan of $S$, which takes $O(1)$ time. Kirsch *et al.* [8] show that a cuckoo hash table with a stash of size $s$ overflows with probability $O(1/n^{s+2})$.

Recently Goodrich and Mitzenmacher [6] showed that one can construct a cuckoo hash table of size $n$ with a stash of size $s$ obliviously using $O(n + s)$ accesses to the data repository and assuming access to private workspace of size $O(n^\nu)$.

### 2.2.2 Oblivious Simulation

We are now ready to describe the oblivious RAM simulation by Goodrich *et. al.* [7]. The memory at the data repository consists of a cache $C$ for $q$ data items, cuckoo hash tables $T_1, \cdots, T_L$ and a stash $S$. (See Figure 2(a).) The size of the cache, $q$, is $O(\log n)$ and stash $S$ has also size $O(\log n)$. Each table $T_i$ has size $2^i q$ and $L$ is the first integer $i$ such that $|T_L| \geq n$, hence $L$ is $O(\log n)$. Each cuckoo hash table $T_i$ is initialized with two hash functions, $h_1^i$ and $h_2^i$. Stash $S$ is shared between all $L$ tables and is large enough to avoid overflows in tables $T_1, T_2, \cdots, T_L$ with high probability [7].

On a request for item $x$, the client executes Algorithm 2. After $q$ requests cache $C$ becomes full and we obliviously move elements from $C$ to $T_1$. The move consists of creating a new cuckoo hash table $T_1$ from the elements in $C$. The next time $C$ becomes full, instead of moving $C$ to $T_1$, we move the items from $C$ and $T_1$ to $T_2$. Similarly, when we are about to move items to $T_i$ for the second time, instead we move all items from $C, T_1, \ldots T_{i-1}$ to the first empty table among $T_i$ through $T_L$. Oblivious construction of cuckoo hash table $T_i$ takes $O(|T_i|)$ accesses to the data repository. Since table $T_i$ is rebuilt every $O(|T_i|)$ requests and eventually every request causes $O(\log n)$ tables to be rebuilt, the amortized overhead is $O(\log n)$ accesses per request.

*Obliviousness:* Scans and writes to cache $C$ and stash $S$ are oblivious since it is done sequentially.

Table $T_i$ is accessed either according to a pseudo-random hash function or in random locations because item $x$ is found in an earlier level of the construction. Cache $C$ and tables $T_j$, $j < i$ are empty when $T_i$

**Algorithm 2** The request phase during oblivious RAM simulation with the $\log n$ hierarchical approach [7].

    $found \leftarrow false$
    scan cache $C$ and stash $S$. if $x$ is found in one of them set $found \leftarrow true$
    **for** each level $i$, $1 \leq i \leq L$ **do**
        if $found$ is $true$ access random locations in $T_i$.
        else access locations $h_1^i(x)$ and $h_2^i(x)$ in $T_i$.
        if $x$ is found set $found \leftarrow true$
    **end for**
    Remove $x$ from $S$ if $x$ was found in $S$. Rewrite $S$.
    Rewrite $C$, adding or replacing data item $x$.

is rebuilt and hence are used to cache items found in $T_i$. Since $T_i$ is emptied and initialized with two new pseudo-random hash functions as soon as levels above it become full, it is never accessed twice for the same item. Hence, from the point of view of the adversary, Bob, accesses to $T_i$ look random.

## 3 De-amortized ORAM Simulation

### 3.1 The Square Root Solution

In this section, we present an oblivious RAM simulation method with $O(\sqrt{n} \log^2 n)$ access overhead in the worst case. This method is based on the square-root approach originally proposed in [5], which has $O(n \log^2 n)$ worst-case access overhead and $O(\sqrt{n} \log^2 n)$ amortized access overhead (see Section 2.1). We present first the simple square-root solution, to demonstrate some of the ideas behind the more efficient technique developed in Section 3.2. Note that only recently an oblivious RAM solution with sublinear worst-case access overhead has been proposed [12].

*Intuition:* The most expensive step of the square-root approach is building a new table, where items and dummy values are ordered using a new pseudo-random permutation. This step is executed every $\sqrt{n}$ requests and takes $O(n \log^2 n)$ accesses. Our idea is to split the accesses for the rebuild into $\sqrt{n}$ batches, each with $O(\sqrt{n} \log^2 n)$ accesses, and to execute each batch after processing a request so that the new table is ready to be used after processing $\sqrt{n}$ requests. We will show how this idea can be implemented while preserving obliviousness and keeping the same asymptotic access overhead and storage overhead as the original method.

*Memory Layout:* We organize the memory on the data repository into five areas. We make use of two buffers, $B_\text{cur}$ and $B_\text{prev}$, each of size $\sqrt{n}$. We also have two tables, $T_\text{cur}$ and $T_\text{next}$, each of size $n + \sqrt{n}$. These tables are built using different pseudo-random permutations on the $n$ data items outsourced by the client and $\sqrt{n}$ dummy values. Finally, we employ a workspace $W$ of size $n + 2\sqrt{n}$ for constructing incrementally the new table, $T_\text{next}$, while the current table, $T_\text{cur}$, and the two buffers, $B_\text{cur}$ and $B_\text{prev}$, are being used to process requests. (See Figure 1(b) for illustration.)

*Initialization:* We split a sequence of requests into *epochs*, where an epoch consists of exactly $\sqrt{n}$ requests. Initially, buffers $B_\text{cur}$ and $B_\text{prev}$ are empty and each of the tables $T_\text{cur}$ and $T_\text{next}$ contains the $n$ items and $\sqrt{n}$ dummy items permuted according to a pseudo-random permutation, where $T_\text{cur}$ uses permutation $\pi_0$ and $T_\text{next}$ uses permutation $\pi_1$.

*Processing an Epoch:* During an epoch, buffer $B_\text{cur}$ caches the $\sqrt{n}$ items being requested in the current epoch while buffer $B_\text{prev}$ caches the $\sqrt{n}$ items that were requested in the previous epoch. Thus, $B_\text{prev}$ is empty during the first epoch. Also, during an epoch, table $T_\text{cur}$ is used for processing requests and workspace $W$ is used to build incrementally a new table, based on a new pseudo-random permutation.

**Algorithm 3** Oblivious RAM simulation with our de-amortized version of the square-root approach.

    Generate pseudo-random permutation function $\pi_{\text{cur}}$
    Initialize table $T_{\text{cur}}$ by storing the $n$ data items and $\sqrt{n}$ dummy items according to permutation $\pi_{\text{cur}}$
    Initialize $W$ with $n$ data items and $\sqrt{n}$ dummy items
    **while** $true$ **do** {process the requests in an epoch}
        Generate pseudo-random permutation function $\pi_{\text{next}}$
        $request\_count \leftarrow 1$
        **while** $true$ **do** {process request for data item $x$}
            $found \leftarrow false$
            Scan all the locations in buffers $B_{\text{cur}}$ and $B_{\text{prev}}$. During the scan, if data item $x$ is found, set $found \leftarrow true$.
            **if** $found$ **then**
                Access location $\pi_{\text{cur}}(n + request\_count)$ in $T_{\text{cur}}$ {containing a dummy item}
            **else**
                Access location $\pi_{\text{cur}}(x)$ in $T_{\text{cur}}$ {containing data item $x$}
            **end if**
            Rewrite $B_{\text{cur}}$, adding or replacing data item $x$
            Execute the next batch of $c\sqrt{n}\log^2 n$ accesses to workspace $W$ to construct table $T_{\text{next}}$ using permutation $\pi_{\text{next}}$
            $request\_count \leftarrow request\_count + 1$
            **if** $request\_count > \sqrt{n}$ **then**
                **break** {end of the epoch}
            **end if**
        **end while**
        Copy the new table from $W$ to $T_{\text{next}}$
        Copy $B_{\text{cur}}$ to $B_{\text{prev}}$
        Copy $T_{\text{cur}}$ and $B_{\text{cur}}$ to $W$
        Empty $B_{\text{cur}}$
        Copy $T_{\text{next}}$ to $T_{\text{cur}}$
        $\pi_{\text{cur}} \leftarrow \pi_{\text{next}}$
    **end while**

*Transitioning to the Next Epoch:* At the end of an epoch, the new table is copied from $W$ to $T_{\text{next}}$. Next, table $T_{\text{cur}}$ and buffer $B_{\text{cur}}$ are copied to $W$. Finally, table $T_{\text{next}}$ is copied to $T_{\text{cur}}$ to accommodate the requests from the next epoch. Also, we overwrite $B_{\text{prev}}$ with items from $B_{\text{cur}}$ and we empty $B_{\text{cur}}$.

*Incremental Table Construction:* The construction of the new table, $T_{\text{next}}$ in workspace $W$ takes as input $T_{\text{cur}}$ and $B_{\text{cur}}$ from the previous epoch. We say that the instance of a data item in $T_{\text{cur}}$ is stale if there is an instance of the same data item in $B_{\text{cur}}$. Using an algorithm from [5], we obliviously filter out the stale instances of the data items and we construct a table for the set consisting of the $n$ data items and $\sqrt{n}$ dummy items, storing them according to newly generated pseudo-random permutation. Since this algorithm performs $O(n \log^2 n)$ accesses to the data repository, we de-amortize it by splitting its sequence of accesses to workspace $W$ into $\sqrt{n}$ batches of $c\sqrt{n}\log^2 n$ accesses each, for some constant $c > 0$. The construction of table $T_{\text{next}}$ starts at the beginning of the epoch and a batch of accesses is executed after processing each request. Hence, the new table $T_{\text{next}}$ is ready by the end of the epoch.

Our oblivious RAM simulation algorithm based on the square-root approach is outlined in Algorithm 3 and its properties are summarized in Theorem 1.

**Theorem 1.** *Our oblivious RAM simulation method based on the square-root approach has $O(\sqrt{n}\log^2 n)$ worst-case access overhead per request, $O(\sqrt{n})$ space overhead at the data repository, and $O(1)$ client memory, where $n$ is the number of data items.*

*Proof.* (SKETCH) The worst-case access overhead of each request is $O(\sqrt{n}\log^2 n)$ since we scan two buffers of size $\sqrt{n}$, access one table entry, and execute $O(\sqrt{n}\log^2 n)$ accesses to perform one batch of the table rebuild. Also, $O(n)$ additional space is used at the server.

We now consider the obliviousness of our method. During each epoch, unique items are accessed in table $T_{\text{cur}}$. Namely, if the requested data item is not found in the buffer, we access it in $T_{\text{cur}}$, else we access a new dummy item in $T_{\text{cur}}$. Moreover, in the beginning of each epoch $T_{\text{cur}}$ is initialized with a new permutation over $n$ items and $\sqrt{n}$ dummy values.

The method is correct since the user is always returned the most up-to-date instance of the requested item: if the requested data item was last requested in the current epoch then it is found in $B_{\text{cur}}$, else if it was last requested in the previous epoch, it is found in $B_{\text{prev}}$, else $T_{\text{cur}}$ has the latest instance. □

*Read/Write Data Repository:* In Algorithm 3 we made an assumption that data repository allows us to manage outsourced memory using *copy* operator. However, it is not required for execution of our de-amortized method and achieving the same worst case overhead of $O(\sqrt{n}\log^2 n)$. We provide only an intuition behind this approach. If the data repository supports only read and write operations one can alternate blocks of memory used for rebuild and for handling requests between the epochs, e.g. during even numbered epochs $B_{\text{cur}}$ is used to cache current requests while during odd epochs it serves as a buffer of requests from the previous epoch.

## 3.2 The $\log n$ Hierarchical Solution

We now describe the de-amortization of the oblivious RAM simulation method presented in [7], which is based on a hierarchical memory layout at the server and has $O(\log n)$ amortized access overhead. The intuition behind the de-amortized version of this method is similar to the one we used in the square-root solution (see Section 3.1): we incrementally rebuild tables while handling requests using previous versions of tables and buffers. Requests are handled using two sets of buffers: one for items requested in the current epoch and the other for items requested in the previous epoch. However, recall that the construction of [7] has $O(\log n)$ buffers implemented as cuckoo hash tables. This complicates our task since now we need to have a copy of each cuckoo hash table. Also due to the dynamic arrangement of the buffers, one buffer spills into the next one and so on, we eventually need to construct $O(\log n)$ cuckoo hash tables during each epoch.

*Memory Layout:* Our memory layout at the data repository is schematically illustrated in Figure 2(b). Extending the oblivious RAM data structure of [7] (see Section 2.2), we employ two caches, $C_{\text{cur}}$ and $C_{\text{prev}}$, of size $q = O(\log n)$, one stash, $S$, of size $O(\log n)$, and $2L - 1$ cuckoo hash tables $T_1, T_1', T_2, T_2', \ldots, T_{L-1}, T_{L-1}', T_L$ where each $T_i$ and $T_i'$ has size $2^i q$ and $L$ is the smallest $i$ such that $2^i q \geq n$. We also keep a workspace $W$ for rebuilding cuckoo hash tables. The workspace stores the last $2^L q$ requested items in a list, $D$. In addition, it contains $L$ work areas for rebuilding cuckoo tables. The $i$-th work area consists of storage space $T_i^W$, of size $O(2^i q)$, for a cuckoo hash table at level $i$ and of overflow space $S_i^W$, of size $O(\log n)$, for the corresponding stash.

*Initialization:* We build $T_L$ as a cuckoo hash table for the $n$ data items and put into stash $S$ any items that did not fit. Both caches $C_{\text{cur}}$ and $C_{\text{prev}}$ and other tables $T_i$ and $T_i'$ (for $i < L$) are empty.

*Processing an Epoch:* In this section an epoch is defined as a sequence of $q$ requests. During an epoch, cache $C_{\text{cur}}$ stores data items last requested in the current epoch and cache $C_{\text{prev}}$ stores data items last requested in the previous epoch. Thus, these caches play roles similar to those of buffers $B_{cur}$ and $B_{prev}$ in Section 3.1. Each request is processed by scanning caches $C_{\text{cur}}$ and $C_{\text{prev}}$, scanning stash $S$, and accessing locations in tables $T_1, T_1', T_2, T_2', \ldots, T_{L-1}, T_{L-1}', T_L$. In addition, a batch of accesses is made to workspace $W$ toward rebuilding its cuckoo tables. The incremental rebuilding process guarantees the completion of a cuckoo table in $T_i^W$ and its stash in $S_i^W$ after $2^{i-1}$ epochs.

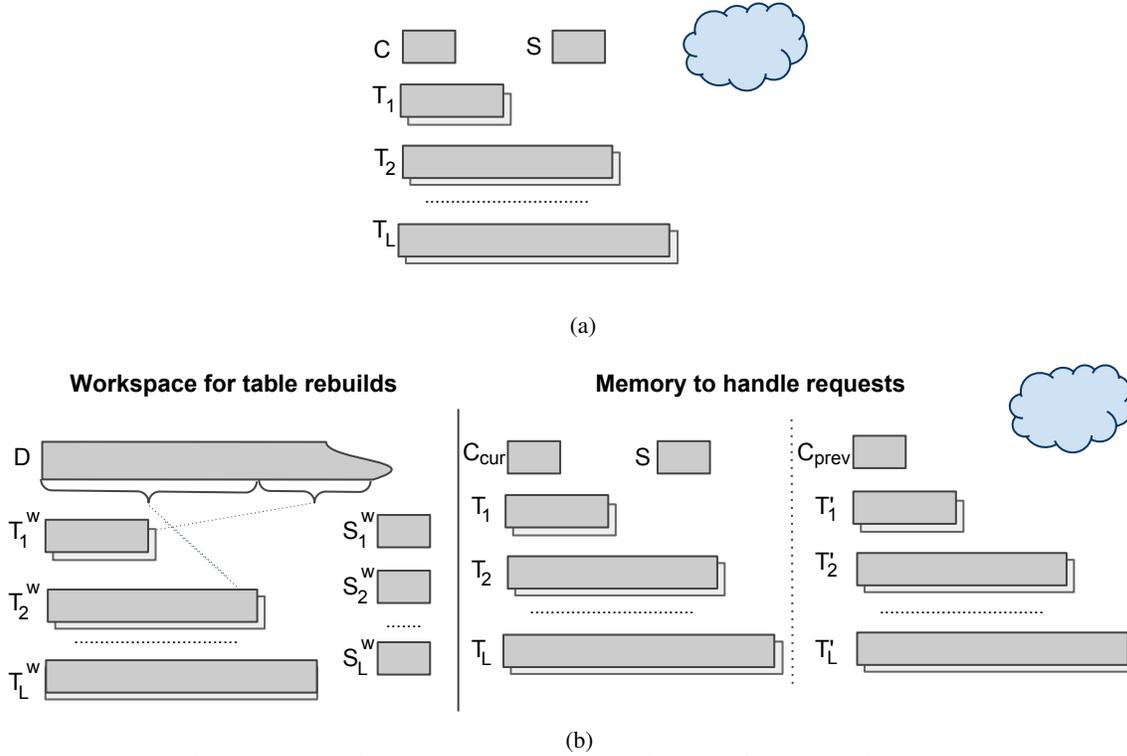

(a)

(b)

Figure 2: Memory layout of the data repository during oblivious RAM simulation: (a) original version of the $\log n$ hierarchical solution [7]; (b) our de-amortized version (Section 3.2).

*Incremental Construction of $L$ Cuckoo Hash Tables:* Recall that a cuckoo hash table $T_i$ of size $2^i q$ can be constructed obliviously using $2^i q$ accesses to the data repository and $O(n^\nu)$ private memory [6]. To help us explain the concurrent oblivious rebuild of $L$ cuckoo hash tables, we introduce a data structure $I$ that stores the set of indices of the cuckoo tables that need to be rebuilt in workspace $W$. Note that $Queue$ depends only on the number of requests made so far hence it can be computed in constant time. $I$ starts empty. After every $2^{i-1}$ epochs index $i$ is added to $I$. When an index $i$ is added to $I$ a sequence of $2^i bq$ accesses is required for a rebuild of $T_i^W$, for some constant $b > 0$. After each request, the client executes $2b$ accesses for each index in $I$ so that the construction of table $T_i^W$ is completed in $2^{i-1}$ epochs. Observe that after the first $2^{i-1}$, epochs index $i$ is always present in $I$. Moreover, after $2^{L-1}$ epochs indices $1, 2, \ldots, L$ are present in $I$ and $I$ does not change from then on. This also means that eventually all $L$ tables are being rebuilt during an epoch. (See Figure 3 for an illustration of the rebuilding process.) To accommodate $L$ concurrent rebuilds, we increase the requirement on the size of client's private memory from $O(n^\nu)$ in [7] to $O(n^\tau)$, for some fixed constants $\tau > \nu$ and $\nu > 0$.

*Transitioning to the Next Epoch:* We append items in $C_{\text{cur}}$ to $D$, the list in workspace $W$ that keeps track of the $2^L q$ previously requested items. We then copy $C_{\text{cur}}$ to $C_{\text{prev}}$ and empty it. Hence $C_{\text{cur}}$ can be used to cache requests during the next epoch. We then check which tables are finished, i.e. $T_i^W$ is finished if the current number of epochs is a multiple of $2^{i-1}$ since $T_i^W$ takes $2^{i-1}$ epochs for a rebuild. Each such table $T_i^W$ is then copied to either $T_i$ or $T_i'$. If $T_i$ and $T_i'$ are both empty or both full $T_i^W$ is copied to $T_i$, stash $S_i^W$ is merged with $S$ and $T_i'$ is cleared. If only $T_i$ is full $T_i^W$ is copied to $T_i'$, stash $S_i^W$ is merged with $S$. If $T_L$, the table from the last level, is finished we clear first $2^L q$ items from $D$ since all these items are now in $T_L$ and no table from earlier levels requires them for a rebuild.

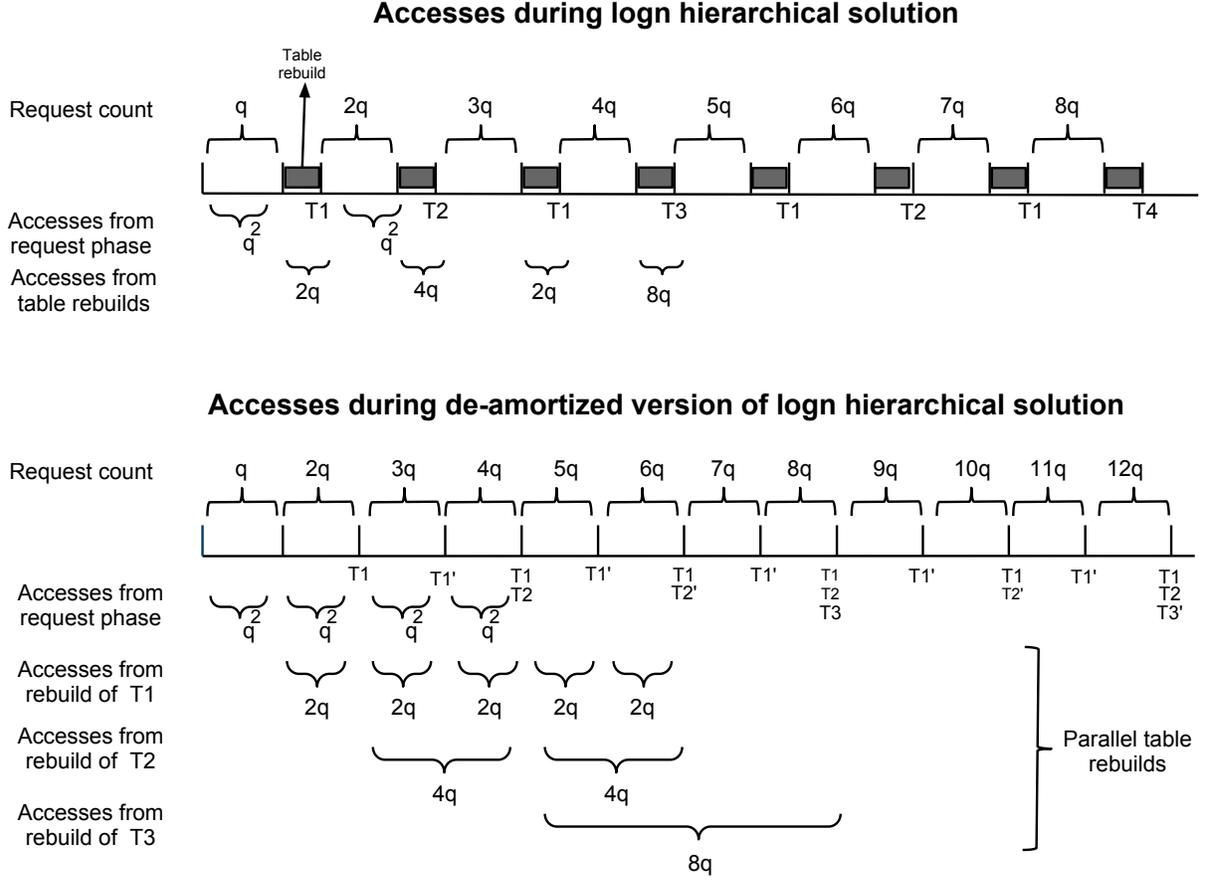

Figure 3: Sequence of accesses during oblivious RAM simulation: (a) original version of the $\log n$ hierarchical solution [7]; (b) our de-amortized version (Section 3.2). The size of the cache is denoted by $q$, which is $O(\log n)$.

*Stash Size:* Goodrich *et al.* [7] show that a single stash of size $O(\log n)$ is enough to avoid overflows in cuckoo hash tables $T_1, \ldots, T_L$ where $T_i$ contains $2^i q$ items. In our construction, we use a single stash $S$ for two collections of cuckoo hash tables. A single stash ensures that if item $x$ happened to not fit into two tables $T_i$ and $T'_j$ then only the most recent copy is present in $S$. One can view stash $S$ as a joint stash between tables $T_1, \ldots, T_L$ and $T'_1, \ldots, T'_{L-1}$. Suppose a stash of size $s \log n$ is used in the construction of [7], where $s > 1$ is a constant. Then we set our stash $S$ to be of size $2s \log n$, where the first $s \log n$ locations are used for tables $T_1, \ldots, T_L$ and the remaining $s \log n$ locations for $T'_1, \ldots, T'_{L-1}$, with the additional constraint that only unique items can be present in $S$. The latter constraint is enforced when we merge stash $S_i^W$ of a new table $T_i^W$ with $S$.

Our oblivious RAM simulation algorithm based on the $\log n$ hierarchical approach is outlined in Algorithm 4 and its properties are summarized in Theorem 2.

**Theorem 2.** *Our oblivious RAM simulation method based on the $\log n$ hierarchical approach has $O(\log n)$ worst-case access overhead per request, $O(n)$ space overhead at the data repository, and $O(n^\tau)$ client memory, where $n$ is the number of data items and $\tau$ is any fixed positive constant.*

*Proof.* (SKETCH) We first show that handling of each request using the above protocol takes $O(\log n)$ accesses. Retrieving a data item takes $O(\log n)$ accesses since three blocks of size $O(\log n)$ are scanned

**Algorithm 4** Oblivious RAM simulation with our de-amortized version of the $\log n$ hierarchical approach.

    Initialize $T_L$ and $S$ with a cuckoo hash table with a stash using $n$ data items.
    $I \leftarrow \{\}$, $request\_count \leftarrow 0$
    **while** $true$ **do**
      **while** $true$ **do** {on request $x$}
        $found \leftarrow false$
        Scan all the locations in caches $C_{\text{cur}}$ and $C_{\text{prev}}$ and stash $S$. During the scan, if data item $x$ is found, $found \leftarrow true$.
        **for** each level $i$, $1 \leq i \leq L$ **do**
          **if** $i \neq L$ and $T'_i$ is not empty
            if $found$ is $true$ access random locations in $T'_i$.
            else access locations $h_1^{i'}(x)$ and $h_2^{i'}(x)$ in $T'_i$.
            if $x$ is found, $found \leftarrow true$.
          **if** $T_i$ is not empty
            if $found$ is $true$ access random locations in $T_i$.
            else access locations $h_1^i(x)$ and $h_2^i(x)$ in $T_i$.
            if $x$ is found, $found \leftarrow true$.
        **end for**
        Rewrite $C_{\text{cur}}$, adding or replacing data item $x$.
        if $x$ is found in stash $S$ remove $x$ from $S$. Rewrite $S$.
        **for** $i \in I$
          Make next $2b$ accesses towards a rebuild of table $T_i^W$
        $request\_count \leftarrow request\_count + 1$
        **if** $request\_count \mod q = 0$ **then**
          {end of the epoch}
          Copy $C_{\text{cur}}$ to $C_{\text{prev}}$ and append it to $D$.
          Empty $C_{\text{cur}}$.
          **for** $i \in sorted\_decr\_order(I)$
            **if** $request\_count \mod 2^{i-1}q = 0$
              **if** $i = L$
                copy $T_i^W$ to $T_L$.
              **else if** $T_i$ and $T'_i$ are both full or both empty
                Empty $T'_i$ and copy $T_i^W$ to $T_i$.
              **else**
                copy $T_i^W$ to $T'_i$.
              Merge $S_i^W$ and $S$.
          **for** each level $i$, $1 \leq i \leq L$
            **if** $request\_count \mod 2^{i-1}q = 0$
              Copy last $2^{i-1}q$ items from $D$ to $T_i^W$
            **if** $i \notin I$
              $I \leftarrow I \cup \{i\}$.
            **if** $i = L$
              empty $D$.
        **end if**
      **end while**
    **end while**

and two accesses are made to $2L$ tables, where $L$ is $O(\log n)$. The batch of accesses for table rebuilding made after each request consists of $2b$ accesses for every table in $I$, where $I$ has at most $L$ indices and $b$ is a constant. The method clearly requires only $O(n)$ space on the data repository. For every rebuild, we use the method of [6] which requires $O(n^\nu)$, $\nu > 0$, of client private memory. Since our method concurrently makes $O(\log n)$ rebuilds $O(n^\tau)$ of private memory is required for $\tau > \nu$.

We now consider the obliviousness of the method. Table rebuilds remain oblivious since they follow a predetermined schedule that depends on $n$ and $request\_count$ and are performed in the same way as in the original ORAM construction in [7].

It remains to show that each request remains oblivious. Accesses to the caches, $C_{\text{cur}}$ and $C_{\text{prev}}$, and the stash, $S$, are oblivious since their memory locations are scanned (read and rewritten entirely) for each request. Since accesses to table $T_i$ depend on whether an item is found in $T'_i$ we first show the obliviousness of access sequence to $T'_i$. Observe that when $T'_i$ is substituted with a new cuckoo hash table $T^W_i$ cache $C_{\text{cur}}$ and all tables $T'_{j<i}$ are empty. Since each table on level $i$ can store up to $2^{i-1}q$ items before it is emptied there is space to remember the following number of requests:

$q + \sum_{j=1}^{i-1} 2^{j-1}q = 2^{i-1}q$.

If an item is not found in previous levels, it is accessed according to pseudo-random hash functions $h_1^{i'}$ and $h_2^{i'}$. Otherwise, $T'_i$ is accessed in random locations. $T'_i$ is cleared as soon as next table for this level, i.e. next $T^W_i$, is ready. This happens $2^{i-1}q$ requests after $T'_i$ was last substituted with a new table. Hence, $T'_i$ is never accessed more than once for the same item.

An access to table $T_i$ follows an access to $T'_i$ and is random if an item is found in earlier tables or in $T'_i$. Note that similarly to $T'_i$ cache $C_{\text{cur}}$ and all tables $T'_{j<i}$ are empty when $T_i$ is ready. Moreover $T'_i$ is empty as well. Hence, there is space to remember the following number of requests:

$q + \sum_{j=1}^{i-1} 2^{j-1}q + 2^{i-1}q = 2^i q$.

However, $T_i$ is replaced with a new table every $2^i q$ requests. Hence, no location is accessed more than once in table $T_i$ as well.

To prove the correctness of the method, we observe that the most current copies of the data items are present in the caches or smaller tables. Moreover, table $T'_i$ contains more recent requests than $T_i$ and stash $S$ contains any items that did not fit in their corresponding tables. When newly constructed tables are moved from $W$ to the memory for handling requests, we merge the stash of larger tables with $S$ first. In this case, if the same item did not fit into more than one table, only the most recent copy is in $S$. Since we first scan the caches, the stash and start accessing tables from smaller levels, with $T'_i$ before $T_i$, our method returns to the user the latest instance of the requested item. □

*Read/Write Data Repository:* Similar to de-amortized version of square root solution from Section 3.1 we can relax the assumption of the interface that data repository provides us. If read and write are the only supported operations we can alternate the blocks of memory used for rebuilds and for handling the requests depending on the epoch count.

## 4 Conclusion

We have presented methods for oblivious RAM simulation with efficient worst-case access overhead. Our methods are based on de-amortizing two known solutions, the square root approach and the $\log n$ hierarchical approach, which have efficient amortized access overhead but $\Omega(n)$ access overhead in the worst case. For each of our methods, the worst-case access overhead per request is asymptotically equal to the amortized access overhead of the solution it is based on. In particular, our $\log n$ hierarchical solution incurs $\log n$ access overhead on every request.

**Acknowledgements**

This research was supported in part by the National Science Foundation under grants 0721491, 0847968, 0915922, 0953071, 0964473, 1011840, and 1012060.